\documentstyle[aps,prl,graphics,preprint]{revtex}

\begin{document}
\draft
\title{Direct comparison between potential landscape and local density of states in a disordered two-dimensional electron system}
\author{M. Morgenstern$^1$, J. Klijn$^1$, Chr. Meyer$^1$, M. Getzlaff$^1$, R. Adelung$^2$, R. A. R\"omer$^3$, K. Rossnagel$^2$, L. Kipp$^2$, M. Skibowski$^2$, and R. Wiesendanger$^1$}
\address{$^1$Institute of Applied Physics, 
Hamburg University, Jungiusstra\ss e 11, D--20355 Hamburg, Germany, $^2$ Institute for Experimental and Applied Physics, Christian-Albrechts-University Kiel,
Leibnizstra\ss e 19, D-24098 Kiel, Germany, $^3$ Institute of Physics, Chemnitz University of Technology, 09107 Chemnitz, Germany}
\date{Submitted 01 February, 2002}
\maketitle

\begin{abstract}
The local density of states (LDOS) of the adsorbate induced two-dimensional electron system (2DES) on n-InAs(110) is studied by low-temperature scanning tunneling spectroscopy.
The LDOS exhibits irregular structures with fluctuation lengths decreasing with increasing energy. 
Fourier transformation reveals that the $k$-values of the unperturbed 2DES dominate the LDOS, but additional lower $k$-values contribute significantly.
To clarify the origin of the additional $k$-space intensity, we measure the potential landscape of the same 2DES area with the help of the tip induced quantum dot. 
This allows to calculate the expected LDOS from the single particle Schr\"odinger equation and to directly compare it with the measured one.  
Reasonable correspondance between calculated and measured LDOS is found.
\end{abstract} 
\vspace{0.5cm}
\narrowtext


Two-dimensional electron systems (2DES) are intensively studied as a paradigmatic case for many-particle systems in disordered potentials \cite{2DES}. 
They exhibit unique properties with respect to their three-dimensional counterparts such as 
weak localization or the quantum Hall effect \cite{And}. Although
many experiments probed their macroscopic properties, 
little is known about the underlying local
density of states (LDOS). In fact, only one study deals with the spatial distribution of the 2DES LDOS, which, however, reveals little quantitative information, 
because neither the exact subband energies nor the disorder potential were known \cite{Yama}. 
Here, we investigate the LDOS of the adsorbate induced 2DES on InAs(110) \cite{FeInAs} by scanning tunneling spec\-tro\-scopy (STS). 
Subband energies are determined by angle-resolved photoelectron spec\-tro\-scopy (ARUPS) and
the 2DES potential landscape is measured using the lowest state of the tip induced quantum dot (QD)  
\cite{QD}. This, for the first time, allows a direct comparison between disorder potential and LDOS. 


The UHV-low temperature STM working at $T=6$ K with spectral resolution in STS down to 0.5 mV is described elsewhere \cite{STM}.
Degenerate n-InAs ($N_{\rm D} = 1.1 \times 10^{16} / $cm$^{3}$) is 
cleaved in-situ at a base pressure of $10^{-8}$ Pa, which leads to a nearly defect free InAs(110) surface
with a Fermi level $E_F=5$ meV above the conduction band minimum.
To induce the 2DES, Fe is deposited on the surface from an
e-beam evaporator. The Fe coverage is determined by counting the Fe-atoms and given with respect
to the unit cell of InAs(110). The cleanliness of the Fe is checked by Auger electron spectroscopy. 
Topographic STM-images are recorded in constant current mode 
with voltage $V$ applied to the sample. The $dI/dV$-curves are measured  by lock-in technique ($f =$ 1.5 kHz, $V_{mod}=$1.8 mV) with fixed tip-surface distance 
stabilized at current $I_{stab}$ and voltage $V_{stab}$. 
The $dI/dV(V)$-images are measured point by point, each moving the voltage to $V$ after stabilizing the tip at $V_{stab}$ and $I_{stab}$. 
The influence of the spatially changing tip-surface distance is checked to be of minor importance and thus neglected \cite{PRL1}.
ARUPS experiments are performed on identically prepared samples using the HONORMI beamline at 
HASYLAB with photon energy h$\nu = 10$ eV and an ASPHERE analyzer. The total energy resolution was 20 meV and the angular resolution
0.25$^\circ$ in one and 0.45$^\circ$ in the other direction \cite{Kiel}. The Fermi level is determined on a clean Ta-foil  with an accuracy of 5 meV. For the
ARUPS measurements the Fe flux is calibrated by a quartz balance.

Since a perturbing influence of the STM tip on the LDOS data cannot be excluded in STS on semiconductors \cite{QD}, we determine the subband energies 
of the 2DES independently by ARUPS. Previous ARUPS measurements revealed the coverage dependence of the adsorbate induced
band shift and a rough estimate of the 2DES subband energies $E_n$ \cite{FeInAs}. 
With the high resolution of the ASPHERE analyzer, individual subband peaks are resolved (Fig.~\ref{Fig1}a, points). A straightforward  
fit of the data with the $E_n$'s as the only fitting parameters \cite{FeInAs} leads to
$E_1 = -105 \pm 5$ meV  and $E_2 = -40 \pm 5$ meV (see lines in Fig.~\ref{Fig1}a). 
The resulting $E_n$'s are additionally validated by measuring the angular dependence of the 2DES peak and fitting the data with the same procedure (not shown).\\ 
Next, we compare the ARUPS data with STS. Fig.~\ref{Fig1}b and d show spatially averaged $dI/dV$-curves
representing the macroscopic average 
of the LDOS: the DOS \cite{interact}. The curves in Fig.~\ref{Fig1}b are measured with the same microtip before and after Fe-deposition. 
Without Fe, two peaks caused
by the tip induced quantum dot (QD) appear \cite{QD}. With Fe, the lower peak shifts to lower energies while the other disappears. The shift of the lower peak is
caused by the adsorbate induced band bending. Indeed, the surface band shift of 300 meV measured by ARUPS requires a peak shift of 
80 meV as evidenced by solving an equivalent of the Poisson-Schr\"odinger equation \cite{FeInAs,QD}. 
The disappearance of the second peak is caused by the reduction of the QD size due to the
screening by the 2DES.\\
Between the QD peak and $E_F$, a rather flat $dI/dV$-intensity with two step-like features at $-108$ mV and $-43$ mV is found, which has to be attributed to the 2DES.
Since the features are located  close to the $E_n$'s determined by ARUPS, 
we identify them with $E_1$ and $E_2$. Additional evidence comes from Fig.~\ref{Fig1}c, a grey scale plot of $dI/dV(V)$ along a
substrate line. The 2DES region exhibits intensity fluctuations with a fluctuation length decreasing abruptly at $E_2$. This result is
straightforwardly explained by the fact that the DOS doubles abruptly at $E_2$. The number of states contributing to the image doubles, and since each state has a different spatial phase,
the fluctuation length decreases.\\
Fig.~\ref{Fig1}d shows another spatially averaged $dI/dV$-curve recorded with a different tip at slightly lower Fe coverage and $E_n$'s marked again. 
The QD states are absent and clear step like structures as expected from a 2DES DOS are visible at the $E_n$'s.
We conclude that the presence of the QD does not change the energies in the step like DOS, but slightly influences the intensity distribution.\\
The presence of the QD provides a unique advantage.  As described elsewhere,
the energy of the lowest QD state follows the electrostatic potential in the center of the QD \cite{QD}. Since the extension of the QD state perpendicular to the surface is the same as
the extension of the 2DES, the QD state directly monitors the local 2DES potential. Indeed, the QD energy fluctuates with position as visible in Fig.~\ref{Fig1}c (curved line along y-axis at QD).
A plot of the QD energy as a function of position is shown in Fig.~\ref{Fig2}a. Four troughs about 20~meV in depth are visible. This is exactly the number of substrate donors on average 
located in such a 2DES area. Moreover, 20~meV is exactly the attractive potential 
of a single donor averaged over the extension of the 2DES. We take both as strong evidence that the QD state indeed maps the 2DES potential.\\
What is the influence of the Fe atoms? An STM-image of a small area of Fig.~\ref{Fig2}a (black square) is given in Fig.~\ref{Fig2}b. It shows several Fe atoms (dark dots), but no 
correspondence between
the Fe positions and the measured potential. This might be surprising, since the adsorbate layer donates electrons to the 2DES and is thus charged \cite{FeInAs}.
Anyway, in the area of Fig.~\ref{Fig2}a only 700 electrons are donated, but 7000 Fe atoms are deposited. Assuming that each Fe atom provides one
electron at $E_F$, an electron density of $1.5 \times 10^{13}$cm$^{-2}$ 
remains in the Fe layer sufficient to screen the positive charge on small length scales. \\ 
Fig.~\ref{Fig2}c shows a more irregular potential obtained at 0.8 \% coverage. It exhibits 
more troughs than expected from the 16 bulk donors. 
Here, the remaining electron density in the Fe layer of 
$2 \times 10^{12}$/cm$^{-2}$ is not sufficient to screen the positive charge of $8 \times 10^{11}$ e/cm$^{-2}$ completely.\\
The measured potential can be used to estimate the mobility of the 2DES \cite{mob}. For Fig.~\ref{Fig2}a it turns out to be $\mu \simeq 5 \times 10^6$ cm$^2/$Vs 
indicating a mean free path in the large ${\rm \mu}$m range. For Fig.~\ref{Fig2}c it is slightly lower. Notice, that the mobility as well as the potential landscape itself is rather similar to 
high-mobility 2DES's \cite{2DESPot}, stressing the relevance of the STS results for transport measurements.\\ 
Next, we discuss the LDOS. Fig.~\ref{Fig3}a$-$g show some of the LDOS images recorded at 2.7 \% coverage in the absence of a QD. The spatial resolution is 5 nm well below the Fermi wave 
length of 23 nm. 
The total intensity in each image would correspond to 30 electronic states, if these states are completely localized in the image area.
But since the scattering length and thus the localization length is probably larger, more states contribute to the LDOS with part of its intensity distribution.
The LDOS images exhibit corrugations
decreasing in length scale with increasing voltage. The corrugation patterns
are rather complicated and do not exhibit the circular structures found in the InAs 3DES \cite{PRL1}.
The corrugation strength defined as the ratio between spatially fluctuating and total $dI/dV$-intensity is $60\pm 5$ \%, i.e. 
much larger than the corrugation strength in the 3DES ($4 \pm 0.5$ \%) \cite{PRL1}. 
Both results reflect the tendency of the 2DES to
weakly localize \cite{And}. Many different scattering paths containing each many scattering events contribute to the LDOS leading to more intricate patterns and
the tendency for localization leads to an increased corrugation.\\    
Fourier transforms (FT's) of the LDOS (insets) reveal the distribution of contributing $k$-values. 
At low voltage a circle is visible in the FT, which at higher voltage is confined by a ring. At even higher voltages ($V>-40$ mV) a second smaller circle appears indicating the 
occupation of the second subband. A plot of the $k$-values corresponding
to the rings is shown in Fig.~\ref{Fig3}h. At low voltages, where the ring is not apparent, the outer diameter of the circle is taken. 
For comparison,
the $E(k)$-dispersion of unperturbed InAs \cite{Merkt} is drawn. 
The correspondence of the dispersion curve  with the data is excellent for the lower subband and slightly worse for the second subband demonstrating that the unperturbed
$k$-values still dominate the spectrum. 
However, additional $k$-space intensity not compatible with the unperturbed dispersion exists in the FT's. It is
strongest within the rings \cite{nonoise}.\\
For 0.8 \% coverage (Fig.~\ref{Fig3}i$-$l), we find the same tendencies as for 2.7 \% coverage.  Here, only one subband is occupied ($E_1=-60$ meV) and
the tip exhibits a QD state. From the QD state, the potential in Fig.~\ref{Fig2}c results and potential and LDOS can be directly compared. This is a crucial result, since effective mass, potential
landscape and electron density completely determine the LDOS, thus all parameters are known. In particular, it allows to show that the 
additional $k$-values contributing to the LDOS are largely caused by the interaction of the electrons with the potential disorder. 
We solve the Schr\"odinger equation for noninteracting particles numerically using 
periodic boundary conditions and the measured disorder potential \cite{Metz,elec}. 
There is no adjustable fitting parameter in the calculation. 
To construct the LDOS, the resulting squared wave functions are weighted corresponding to the energy resolution of the experiment.
The resulting LDOS for a particular energy is shown in Fig.~\ref{Fig4}a in comparison with the measured LDOS in Fig.~\ref{Fig4}b. 
The correspondence is reasonable, i.e. several features as the central ring structure or other smaller structures marked by arrows appear in both images.  
The FT's (insets) and the intensity 
distributions of the LDOS (Fig.~\ref{Fig4}c) even show nearly perfect agreement. We found similar results at the other energies and
conclude that the potential landscape indeed largely determines the LDOS by mixing different $k$-states \cite{tbp}. Remaining discrepancies between measurement and calculation
may be either caused by scattering centers outside the measured region or by electron-electron interactions. However, a study of these effects
is behind the scope of this paper.

In summary, we presented an experimental method to determine the potential landscape and the LDOS of the same disordered 2DES area. 
This is a decisive prerequisite for detailed studies of
the 2DES LDOS under different conditions. The results obtained here are successfully interpreted in terms
of mixing of different $k$-states by the inhomogeneous potential landscape as evidenced by comparing the expected
LDOS calculated from the potential landscape and the measured one. 
 
We thank O. Seifarth and D. Haude for assistance with the measurements. 
Financial support from Wi 1277/15-1 and Gra\-du\-ier\-ten\-kol\-legs ``Phy\-sik na\-no\-struk\-tu\-rier\-ter 
Fest\-k\"or\-per", ``Spektroskopie lokalisierter, atomarer Systeme" of the DFG and the BMBF project 05 KS1FKB is gratefully 
acknowledged.
\narrowtext
\begin{figure}
\caption{a.) ARUPS-spectrum of 4.5 \% Fe/n-InAs(110), h$\nu=10$ eV (points) compared
with fits for different subband energies $E_1$, $E_2$ as indicated (lines). Only the central curve fits the data. b.) Spatially averaged $dI/dV(V)$-curves of n-InAs(110) (lower curve)
and 4.5 \% Fe/n-InAs(110) (upper curve); both curves taken with the same tip, $V_{stab}=100$ mV, $I_{stab}=500$ pA; peaks of the
tip induced quantum dot (QD) and $E_1$, $E_2$ of the 2DES determined by ARUPS as well as the 3DES are indicated. c.) Greyscale plot of $dI/dV(V)$-intensity as a function of position 
along a scan line, 
$V_{stab}=100$ mV, $I_{stab}=500$ pA; sample and tip as in upper curve of b.); $E_1$, $E_2$ and QD peak are indicated. d.) 
Spatially averaged $dI/dV(V)$-curve of 2.7 \% Fe/n-InAs(110), $V_{stab}=100$ mV, $I_{stab}=300$ pA; note the absence of QD peaks.}
\label{Fig1}
\end{figure}
\narrowtext
\begin{figure}
\caption{a.) Potential landscape as determined from laterally fluctuating peak voltage of the lowest-energy QD state, 4.5 \% Fe/n-InAs(110).
b.) Constant-current image of the area marked in a.), $V=100$ mV, $I=50$ pA; dark spots are Fe-atoms. c.) Potential landscape
at 0.8 \% Fe/n-InAs(110). Both potential images cover a potential range of 20 mV.} 
\label{Fig2}
\end{figure}
\narrowtext
\begin{figure}
\caption{a$-$g.) $dI/dV$-images (LDOS-images) of 2.7 \% Fe/n-InAs(110) recorded at different $V$ as indicated; 
$V_{stab}=100$ mV, $I_{stab}=300$ pA; the bright spikes in the images are the Fe-atoms. Insets: 
Fourier transformations (FT) of $dI/dV$-images.
h.) dominating $k$-values corresponding to rings in FT's in comparison with dispersion curve of unperturbed InAs (lines) {\protect \cite{Merkt}}. i.)$-$l.) Same as a.)$-$h.) but for
0.8 \% Fe/n-InAs(110); investigated surface area belongs to the potential in Fig.{\protect \ref{Fig2}c}} 
\label{Fig3}
\end{figure}
\narrowtext
\begin{figure}
\caption{a.) LDOS calculated from the potential landscape in {\protect Fig.~\ref{Fig2}c} {\protect \cite{Metz}}; $E=-50$ meV. b.) Normalized $dI/dV$-image
of the same area; $V=-50$ mV, $V_{stab}=100$ mV, $I_{stab}=300$ pA. Insets are FT's. Dots mark identical sample positions as deduced from constant current images.
c.) Intensity distribution of the LDOS in a.) and b.); for the sake of comparison the experimental curve is stretched by 5 \%.} 
\label{Fig4}
\end{figure}
\newpage
\vspace*{3.5cm}
\centerline{
\resizebox{\textwidth}{!}{
\includegraphics{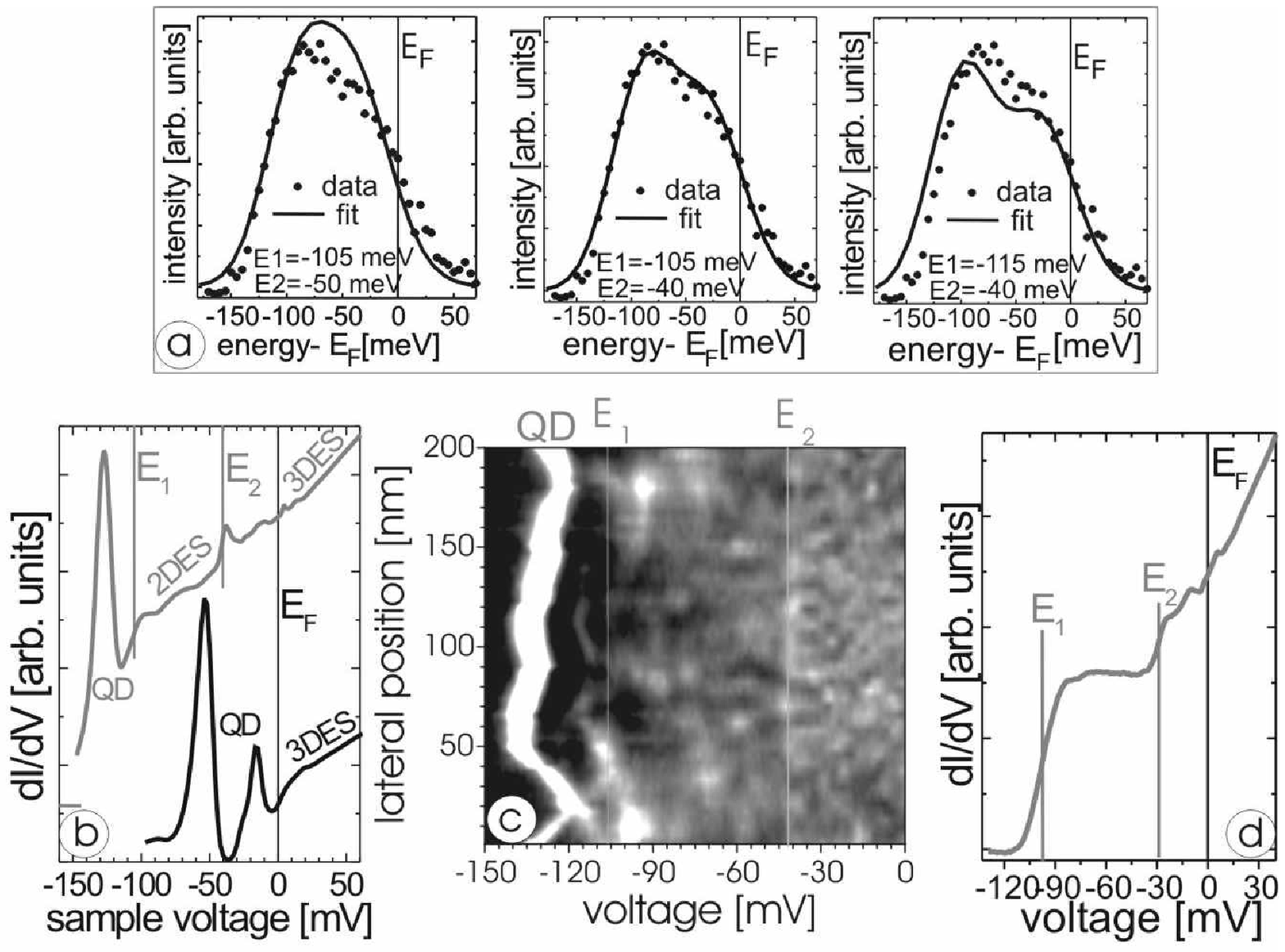}}
}
\vspace{5.7cm}
\centerline{\Huge Fig.~1}
\newpage
\vspace*{4.5cm}
\centerline{
\resizebox{\textwidth}{!}{
\includegraphics{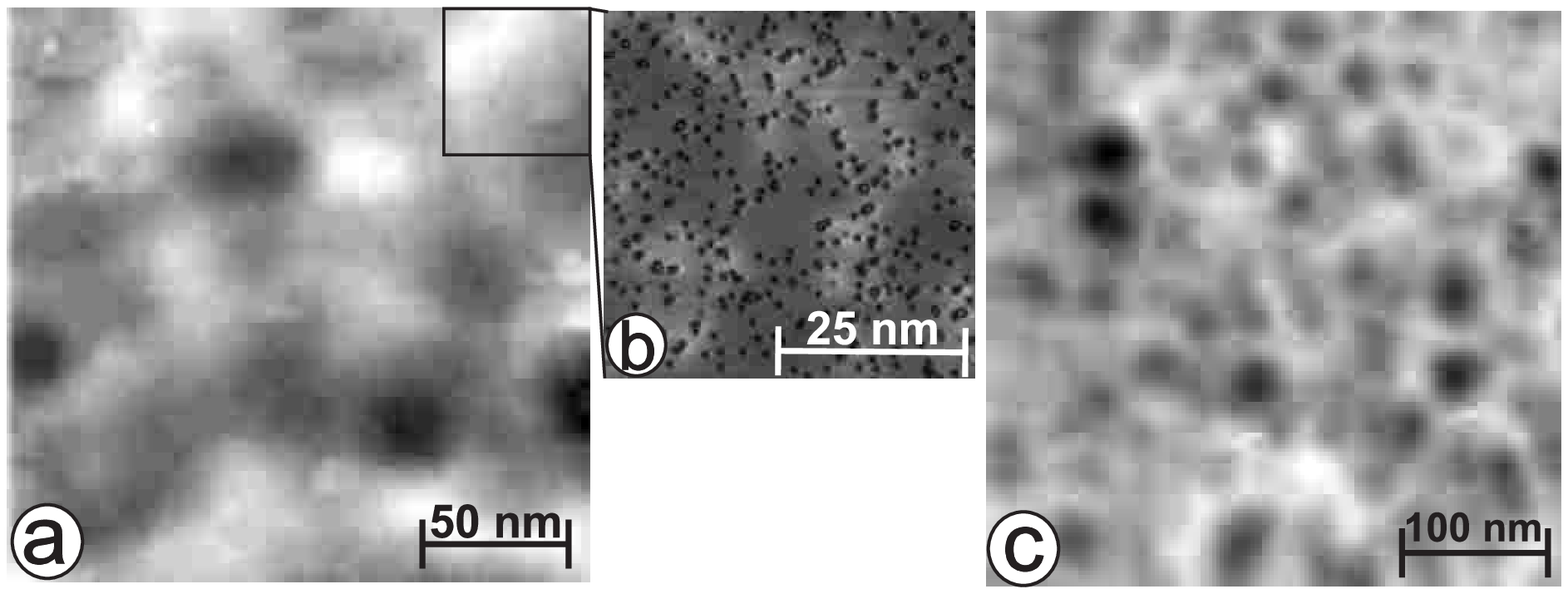}}
}
\vspace{10.8cm}
\centerline{\Huge Fig.~2}
\newpage
\vspace*{4.5cm}
\centerline{
\resizebox{\textwidth}{!}{
\includegraphics{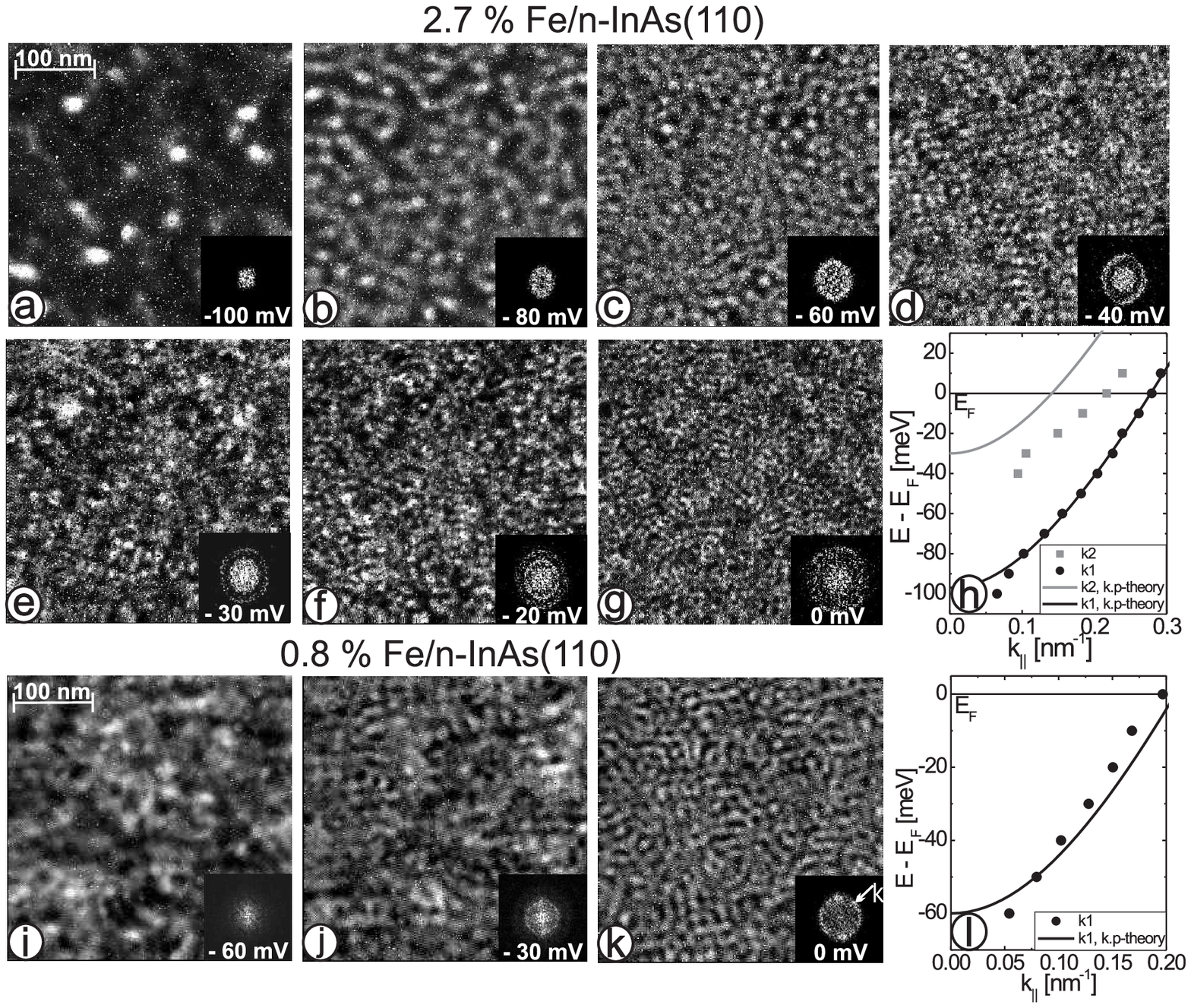}}
}
\vspace{3cm}
\centerline{\Huge Fig.~3}
\newpage
\vspace*{4.5cm}
\centerline{
\resizebox{\textwidth}{!}{
\includegraphics{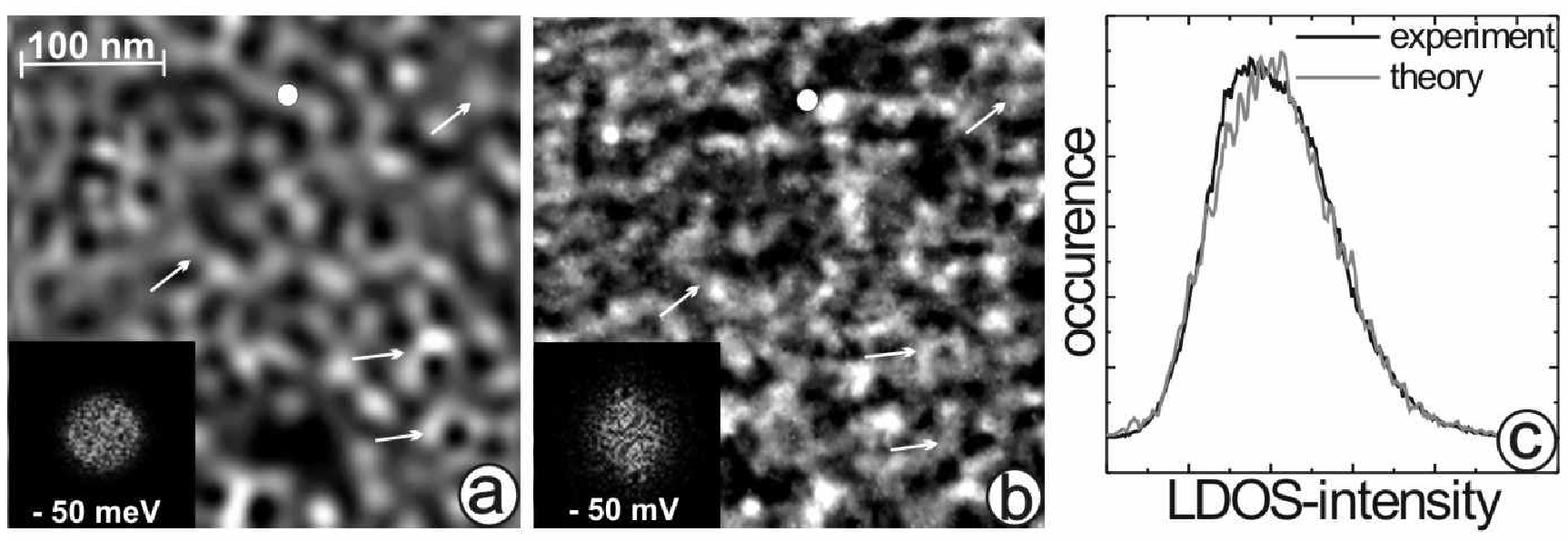}}
}
\vspace{10.4cm}
\centerline{\Huge Fig.~4}

\begin{references}
\bibitem{interact}
LDOS and $dI/dV$ coincide as long as $V$ is small {\protect \cite{QD}}
and electron-electron interactions are negligible.
\bibitem{nonoise}
The experimental noise is considerably smaller than the intensities depicted in the FT's.
\bibitem{elec}
The contributing energy scales are roughly given by the average kinetic energy 
$E_{kin}=30$ meV, the $\sigma$-value of the potential fluctuations $\sigma (V)= 5$ meV and the electron-electron repulsion of regularly distributed point charges 
$E_{Coul}=5$ meV. This indicates that away from $E_F$ electron-electron interactions might be neglected.
\end{references}
\end{document}